\documentclass[amsmath,amssymb,twocolumn,showpacs]{revtex4}

\usepackage{bm}
\usepackage[final]{graphicx}
\usepackage{epsfig}
\usepackage[thinspace]{SIunits}
\usepackage{color}
\usepackage{amsmath}
\usepackage{ulem}

\usepackage{lipsum}
\usepackage{graphicx}

\begin{document}


\title{Electrodynamic response of type II Weyl semimetals}
\date{\today}

\author{M. Chinotti$^+$}
\affiliation{Laboratorium f\"ur Festk\"orperphysik, ETH - Z\"urich, 8093 Z\"urich, Switzerland}

\author{A. Pal$^+$}
\affiliation{Laboratorium f\"ur Festk\"orperphysik, ETH - Z\"urich, 8093 Z\"urich, Switzerland}

\author{W.J. Ren}
\affiliation{Condensed Matter Physics and Materials Science Department, Brookhaven National Laboratory, Upton NY 11973, USA}
\affiliation{Shenyang National Laboratory for Materials Science, Institute of Metal Research, Chinese Academy of Sciences, Shenyang 110016, China}

\author{C. Petrovic}
\affiliation{Condensed Matter Physics and Materials Science Department, Brookhaven National Laboratory, Upton NY 11973, USA}

\author{L. Degiorgi$^\ast$}
\affiliation{Laboratorium f\"ur Festk\"orperphysik, ETH - Z\"urich, 8093 Z\"urich, Switzerland}



\begin{abstract}
Weyl fermions play a major role in quantum field theory but have been quite elusive as fundamental particles. Materials based on quasi two-dimensional bismuth layers were recently designed and provide an arena for the study of the interplay between anisotropic Dirac fermions, magnetism and structural changes, allowing the formation of Weyl fermions in condensed matter. Here, we perform an optical investigation of YbMnBi$_2$, a representative type II Weyl semimetal, and contrast its excitation spectrum with the optical response of the more conventional semimetal EuMnBi$_2$. Our comparative study allows us disentangling the optical fingerprints of type II Weyl fermions, but also challenge the present theoretical understanding of their electrodynamic response.
\end{abstract}

\maketitle

\newpage
Since the discovery of Dirac states in a wide range of materials spanning novel superconductors \cite{Richard2010}, graphene \cite{Novoselov2005} as well as topological insulators \cite{Kane2010}, a great deal of effort has been devoted to observe other types of elementary particles in condensed matter, like Majorana and Weyl fermions. The latter type of fermions may be understood as a pair of particles characterized by opposite chirality, as derived by the massless solution of the Dirac equation \cite{Weyl1929}. In order to observe Weyl fermions in three dimensional condensed matter systems, the requirement of a crossing of two non-degenerate bands, containing relativistic and massless states which just touch in a single point, must be satisfied \cite{Wan2011,Xu2011}. Indeed, a split of the doubly degenerate Dirac point into a pair of Weyl nodes can be induced in the presence of broken time reversal or space inversion symmetry \cite{Burkov2011}.

Broken space inversion symmetry at a single Dirac cone leads to two Weyl nodes hallmarked by the same momentum but shifted by the same amount in opposite directions along the energy axis relatively to each other, while broken time-reversal symmetry results in Weyl nodes at the same energy but separated in momentum space \cite{Wan2011}. Evidences for Weyl fermions with broken space inversion symmetry have been reported in so-called non-centrosymmetric, transition metal monopnictide crystals, like TaAs and TaP \cite{Huang2015,Xu2015}. Moreover, Soluyanov et al. made the proposal for the existence of two types of Weyl semimetal; the standard type I with point-like Fermi surface and the previously overlooked type II arising at the contact of the Fermi level along the line boundaries of electron and hole pockets \cite{Soluyanov2015}. A type II Weyl semimetallic state has been predicted \cite{Z_Wang2016} and observed in MoTe$_2$ \cite{Huang2016}. 

The quasi two-dimensional bismuth layers $A$MnBi$_2$ ($A$ = alkaline as well as rare earth atom) lately advanced as a suitable playground for the investigations of such emergent low-energy quasiparticle excitations \cite{Park2011,Wang2011,Wang2012,May2014}. These non-centrosymmetric and magnetic materials host strong spin-orbital interaction and their two-dimensional network of Bi atoms guarantees Dirac massless dispersions. Furthermore, the broken time-reversal symmetry lifts the degeneracy at the Dirac cones. Density functional theory calculations, supported by tight binding method, on SrMnBi$_2$ and CaMnBi$_2$ first reveal the presence of an anisotropic Dirac cone in the Sr compound and a band crossing occurring along a continuous line in momentum space in the Ca compound \cite{Lee2013}. Most recently, EuMnBi$_2$ and YbMnBi$_2$, isostructural to the Sr and Ca composition respectively, were thoroughly scrutinized with respect to the electronic properties of Weyl fermions \cite{Borisenko2016}. 
Supported by high precision ARPES investigations, Borisenko et al. show that YbMnBi$_2$ (but not EuMnBi$_2$) is a genuine Weyl semimetal of type II. Furthermore, an additional fingerprint was clearly revealed in YbMnBi$_2$, consisting of surface Fermi arcs, as non-trivial surface states connecting the Weyl points \cite{Borisenko2016}.

A variety of representative Dirac and Weyl semimetals, like for instance pyrochlore iridates, quasicrystals and transition metal monopnictides, were recently addressed from the perspective of their optical response \cite{Chen2015,Sushkov2015,Timusk2013,Xu2016,Neubauer2016}. Here, we are triggered by the opportunity to exploit YbMnBi$_2$ and EuMnBi$_2$ as an arena in order to explore the optical response and chase the related fingerprints of a type II Weyl semimetal (i.e., in the Yb-based material) in contrast to its semimetal counterpart (i.e., the Eu compound). Our optical experiment provides evidence for two intervals with a linear frequency dependence of the real part ($\sigma_1(\omega)$) of the optical conductivity in the Yb material, with the slope of the low-energy larger than the one of the high-energy interval. Both linear frequency dependences of $\sigma_1(\omega)$ extrapolates to zero conductivity at the origin of the frequency axis. In the Eu compound only one linear frequency dependence can be clearly identified at high frequencies, cutting the frequency axis at a finite value though, indicating its gapped nature. Our results broadly agree with recent predictions of the optical response in Weyl semimetals \cite{Ashby2014,Carbotte2016_1,Carbotte2016_2} but equally put novel constraints for future theoretical activities. 

Our EuMnBi$_2$ and YbMnBi$_2$ single and well characterized crystals were grown after the procedure described in Ref. \onlinecite{May2014} and \onlinecite{Wang2016}, leading to specimens with shiny surface of typical size 2x2 mm$^2$. We collect reflectivity spectra ($R(\omega)$) from the far-infrared up to the ultraviolet as a function of temperature \cite{comment_Fermi_arc}. This is the prerequisite in order to perform reliable Kramers-Kronig transformation, giving access to all optical functions. We thus achieve the optical conductivity on which we focus here our attention. To this end, standard and well established extrapolation procedures are applied at low as well as high frequencies. In the dc limit (i.e., $\omega\rightarrow$0) we use the Hagen-Rubens (HR) extrapolation of $R(\omega)$ ($R(\omega)=1-2\sqrt{\frac{\omega}{\sigma_{dc}}}$) with dc conductivity in agreement with the transport values \cite{Wang2016}. Further details on the measurements, technique and analysis tools can be found in Ref. \onlinecite{grunerbook}.

\begin{figure}[!htb]
\center
\includegraphics[width=7cm]{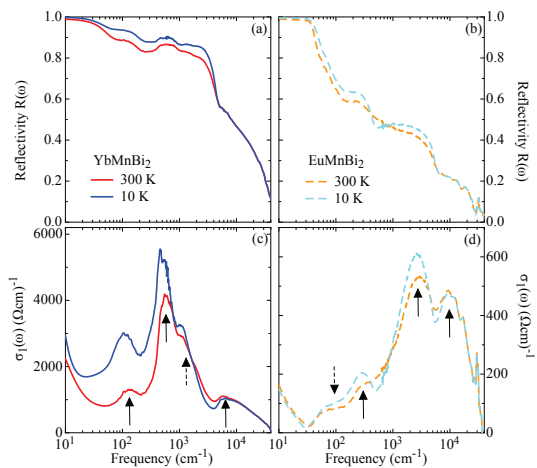}
\caption{(color online) (a) and (b) Reflectivity spectra, and (c) and (d) real part $\sigma_1(\omega)$ of the optical conductivity at 300 and 10 K of YbMnBi$_2$ and EuMnBi$_2$, respectively. Please note the logarithmic energy scale in all panels. The arrows emphasize the pronounced absorptions (plain) and shoulders (dashed) in $\sigma_1(\omega)$ (see text).}
\label{reflectivity_sigma1}
\end{figure}

\begin{figure*}[!htb]
\center
\includegraphics[width=15cm]{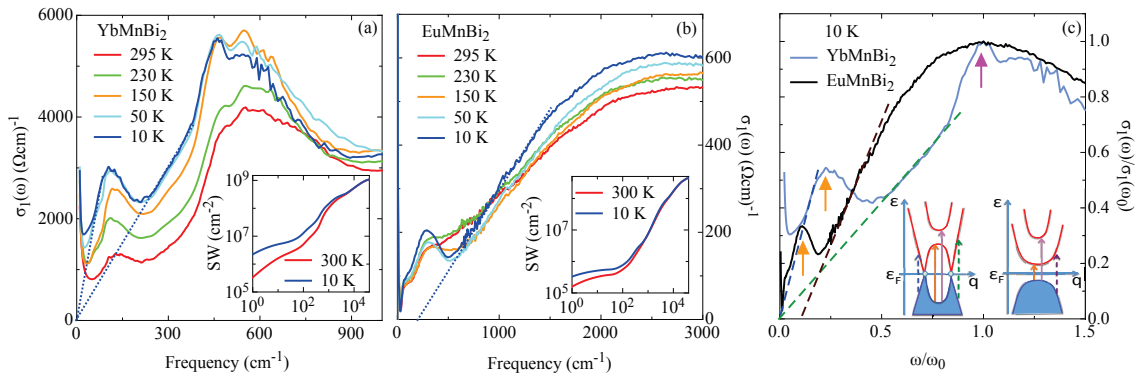}
\caption{(color online) (a) and (b) Temperature dependence of $\sigma_1(\omega)$ in YbMnBi$_2$ and EuMnBi$_2$, respectively, in the energy interval relevant for the present discussion. The insets display the integrated spectral weight of both compounds at 300 and 10 K with double logarithmic scales. The thin dotted lines at 10 K indicate the extrapolation to zero conductance from the quasi-linear frequency dependence in $\sigma_1(\omega)$. The cut with the frequency axis occurs at its origin only in the Yb-compound. (c) Normalized optical conductivity at 10 K with $\omega_0$ = 500 and 2500 cm$^{-1}$ in YbMnBi$_2$ and EuMnBi$_2$, respectively. The dashed lines and plain arrows emphasize the linear frequency dependence of $\sigma_1(\omega)$ in appropriate energy intervals (see text) and the relevant excitations for both compounds, respectively. We use for plain arrows and dashed lines in the main panel, the same color code as for the plain and dashed arrows in both insets, in order to emphasize their relationships. The left inset schematically shows the band structure close to one pair of Weyl nodes (appropriate for the Yb material), with the plain arrows indicating the transitions leading to the van Hove singularities in $\sigma_1(\omega)$ and the dashed ones for the transitions between states with linear dispersion leading to $\sigma_1(\omega)\sim\omega$. The right inset displays the situation of a gapped semimetal (appropriate for the Eu compound) with direct transitions (plain arrows) and transitions between states with vestige of the linear dispersion so that also $\sigma_1(\omega)\sim\omega$ (dashed arrow) \cite{Carbotte2016_2}.}
\label{sigma_1_norm}
\end{figure*}

First of all, we introduce in Fig. \ref{reflectivity_sigma1}(a) and (b) the measured $R(\omega)$ spectra at 10 and 300 K for both compounds. While $R(\omega)$ in the Yb material increases in a rather metallic-like fashion upon decreasing the frequency, with a plasma edge feature below 5000 cm$^{-1}$, $R(\omega)$ of the Eu compound is very much overdamped, merging only at the lower frequency limit of our spectrometer into the metallic HR extrapolation. Nonetheless, there is a quite remarkable temperature dependence, which we will further elaborate below. Before going any further in the data presentation, it is worth pointing out that in the broken space inversion symmetry compound TaAs, $R(\omega)$ gets depleted in the far infrared at low temperatures \cite{Xu2016}, in a quite opposite manner than for our findings. In fact, a sharp plasma edge in $R(\omega)$ of TaAs only develops below about 100 cm$^{-1}$ at 5 K.

The corresponding $\sigma_1(\omega)$ is shown in panels (c) and (d) of Fig. \ref{reflectivity_sigma1}. $\sigma_1(\omega)$ in the Eu material is significantly lower than in the Yb compound. By first inspecting the spectra at low frequencies, we recognize a metallic contribution merging into a Drude-like resonance well below 100 cm$^{-1}$ at all temperatures, in broad agreement with the dc transport properties of both materials \cite{Wang2016}. This bears testimony to a rather small scattering rate (i.e., width of the Drude resonance), so that the metallic component is quite narrow, falling into the spectral range dominated by the HR extrapolation of the measured $R(\omega)$. A narrowing of the Drude resonance has been also encountered in TaAs with decreasing temperature \cite{Xu2016}. The overall metallic contribution is however rather weak, particularly in the Eu compound, which suggests small Drude weight and is quite typical for a semimetallic behavior with moderate-to-low concentration of itinerant charge carriers. This is consistent with the rather fuzzy signature of the Fermi surface, as measured with ARPES technique \cite{Borisenko2016}. Because of the narrow nature of the Drude resonance in $\sigma_1(\omega)$, it is not realistic to artificially disentangle the related temperature dependence of the Drude parameters, plasma frequency and scattering rate.

Three broad peaks at $\sim$ 100, 500 and 6000 cm$^{-1}$ and at $\sim$ 300, 2500 and 10000 cm$^{-1}$ dominate $\sigma_1(\omega)$ in Yb and Eu compound, respectively (see plain arrows in Fig. \ref{reflectivity_sigma1}(c) and (d)). Since the absorptions at 100 and 300 cm$^{-1}$ in the Yb and Eu material, respectively, are rather broad and not compatible with phonon modes, which would generally sharpen with decreasing temperature, we ascribe all three absorptions, shaping the optical conductivity of both compounds, to characteristic electronic transitions \cite{comment_bandstructure1}. We further note the presence of shoulders at the low-frequency side of the peak centered at 300 cm$^{-1}$ in the Eu material, as well as at $\sim$ 1500 cm$^{-1}$ in the Yb compound (see dashed arrows in Fig. \ref{reflectivity_sigma1}(c) and (d)).  

Figures \ref{sigma_1_norm}(a) and (b) emphasize $\sigma_1(\omega)$ at selected temperatures in the spectral range of relevance for the following discussion. The important temperature dependence of $\sigma_1(\omega)$ implies a reshuffling of spectral weight. The insets of panels (a) and (b) in Fig. \ref{sigma_1_norm} deploy the integrated spectral weight ($SW(T,\omega_c) = \int_0^{\omega_c}\sigma_1(\omega,T)d\omega$, $\omega_c$ being a cut-off frequency) at 10 and 300 K in both compounds. It is clearly seen that the spectral weight is fully conserved when integrating $\sigma_1(\omega)$ at any temperatures up to $\omega_c \sim$ 2 eV, thus satisfying the optical sum rules beyond this energy \cite{grunerbook}. There is a piling up of spectral weight at infrared frequencies upon decreasing temperature, which is more substantial in the Yb than in the Eu material and can not be solely ascribed to the Drude narrowing. The full recovery of $SW$ at energies of about 2 eV might suggest that its accumulation at low frequencies upon lowering the temperature mainly results from its progressive shift from the near and mid infrared spectral ranges. 

In order to emphasize the comparison among the excitation spectra of both compounds, we normalize $\sigma_1(\omega)$ at 10 K by the value $\sigma_1(\omega_0)$ at the peak frequency $\omega_0$ = 500 and 2500 cm$^{-1}$ for the Yb and Eu material, respectively, as shown in Fig. \ref{sigma_1_norm}(c). We anticipate at this stage that the low-frequency side of the peaks at 100 and 500 cm$^{-1}$ in the Yb compound can be fairly well approximated with a linear frequency dependence of $\sigma_1(\omega)$, a major fingerprint of this Weyl semimetal (dashed lines in Fig. \ref{sigma_1_norm}(c) as well as thin dotted lines in Fig. \ref{sigma_1_norm}(a)). This is true at all temperatures, although the energy interval of the linearity in $\sigma_1(\omega)$ gets larger at low temperatures. In the Eu compound, on the other hand, a linear frequency dependence of $\sigma_1(\omega)$ is only guessed at the low-frequency side of the strongest peak at 2500 cm$^{-1}$ and most clearly at low temperatures (dashed line in Fig. \ref{sigma_1_norm}(c) as well as thin dotted line in Fig. \ref{sigma_1_norm}(b)). The shoulder on the low-frequency side of the peak at 300 cm$^{-1}$ in the Eu compound (Fig. \ref{reflectivity_sigma1}(d)), prior the sharp onset of the metallic Drude resonance, impedes to single out a robust linearity of $\sigma_1(\omega)$ at such low frequencies, even over a tiny interval. 

Band structure calculations can set the stage for the genesis of Weyl nodes. When spin-orbital coupling is neglected, the expected linear dispersions, generated almost exclusively by Bi$_2$ $p$ states, cross near or at the Fermi level \cite{Lee2013,Borisenko2016}. Such Dirac-like crossings are essential ingredients towards the realization of Weyl nodes and their presence is more favorable in the staggered structure of the Yb atoms than in the coincident one of the Eu atoms relative to the Bi layers. From fully relativistic band structure calculations, taking into account the Mn $d$ electrons magnetism (i.e., antiferromagnetic ordering) and spin-orbital coupling, it can be evinced that two bands due to Mn states and Bi $p$ states contribute to the formation of the Fermi surface \cite{Borisenko2016}. The spin-orbital and exchange interactions lead to 'massive' Dirac states in both materials, so that gaps open in most parts of the Brillouin zone (BZ) and are larger in the Eu than in the Yb compound. Furthermore, the magnetism associated with the Eu element more strongly shifts apart the electronic bands. This is overall consistent with the observation that the optical excitations in $\sigma_1(\omega)$ occur at lower energies in YbMnBi$_2$ than in EuMnBi$_2$. While EuMnBi$_2$ is a semimetal with incipient gapped Dirac cones, it can be shown that the degeneracy at the Dirac points in YbMnBi$_2$ can be lifted by additionally breaking the time reversal symmetry with a canted antiferromagnetic order \cite{comment_magnetism} of the Mn atoms, thus inducing the formation of Weyl nodes. By symmetry and as confirmed by ARPES investigation, there are four pairs of Weyl nodes (so 8 points in total) \cite{Borisenko2016}.

Given the generic features of the electronic band structure, we now turn our attention to the predicted optical conductivity \cite{Carbotte2016_2} in the case of two non-degenerate Dirac cones shifted in reciprocal space by a finite reciprocal q-vector. For simplicity, the Fermi level is pinned to the location of the nodes. Beyond the intraband (Drude) contribution (which might depend very much from disorder and exact location of the Fermi level with respect to the Weyl nodes) the characteristic interband feature in $\sigma_1(\omega)$ should be a kink, generated by the largest absorption between the two paired cones (orange plain arrow in the left inset of Fig. \ref{sigma_1_norm}(c)). This van Hove singularity breaks the expected linear frequency dependence of $\sigma_1(\omega)$ for a single Dirac cone in two quasi-linear parts with variable slopes (associated with transitions connecting states with linear dispersion at the Dirac cones, see blue and green dashed arrows in the left inset of Fig. \ref{sigma_1_norm}(c)). The exact behavior of the interband contribution to $\sigma_1(\omega)$ results from the balance between two parameters: a so-called mass term ($m$) related to the spin-orbital coupling and a magnetic Zeeman-like term ($b$) necessary to break time reversal symmetry. The resonance frequency of the van Hove singularity depends from the relative ratio $m/b$ \cite{Carbotte2016_2}, so that for increasing $m/b$ from 0 to 1 (where the phase boundary to a gapped semimetal is encountered) the location of the kink in $\sigma_1(\omega)$ moves toward low frequencies. For $m/b \rightarrow$ 1, the two quasi-linear regions in $\sigma_1(\omega)$ are characterized by different slopes, so that the slope of the low-frequency response below the kink increases, while the photon-energy range over which it applies decreases. Above the kink, $\sigma_1(\omega)$ rises with a reduced slope with respect to the situation of the massless Dirac cone and the extrapolation of the linear $\sigma_1(\omega)$ to zero frequency cuts the conductivity axis at a finite positive value \cite{Carbotte2016_2}. For $m/b \rightarrow$ 0, $\sigma_1(\omega)$ is supposed to increase linearly with the same slope as for a single Dirac cone on both sides of the kink and the overall linear frequency dependence of $\sigma_1(\omega)$ extrapolates to zero conductance at the origin of the frequency axis. A second van Hove singularity may also occur because of transitions between states due to the lifted degeneracy of the original Dirac cones (pink plain arrow in left inset of Fig. \ref{sigma_1_norm}(c)). For $m/b >$ 1 (i.e., the case for a gapped semimetal), one may expect excitations between gapped and massive Dirac states with lifted degeneracy, intercalated by a quasi-linear frequency dependence of $\sigma_1(\omega)$ reminiscent of the incipient Dirac cones (orange and pink plain as well as brown dashed arrows in right inset of Fig. \ref{sigma_1_norm}(c), respectively) \cite{Carbotte2016_2}.

We propose to identify the broad absorptions at 100 and 500 cm$^{-1}$ in YbMnBi$_2$ with the predicted van Hove singularities (orange and pink plain arrows in Fig. \ref{sigma_1_norm}(c)). Since both linear frequency dependences of $\sigma_1(\omega)$ (blue and green dashed lines and arrows in Fig \ref{sigma_1_norm}(c)), preceding the van Hove singularities, cross zero on the conductivity axis at the frequency origin (Fig. \ref{sigma_1_norm}(a) and (c)) \cite{comment}, we tend to believe that the magnetic term is sizable with respect to the spin-orbital contribution. Nonetheless, it remains to be explained, in contrast to the theoretical predictions \cite{Carbotte2016_2}, why the slopes differ so much among them instead to converge on one single slope as in the case for a Dirac cone. The feature with shoulder around 300 cm$^{-1}$ and the strongest absorption at 2500 cm$^{-1}$ in EuMnBi$_2$ identify the excitations between the gapped Dirac cones, with lifted degeneracy (orange and pink plain arrows in Fig. \ref{sigma_1_norm}(c)). Moreover, the barely perceivable linear frequency dependence of $\sigma_1(\omega)$ (brown dashed arrow and line in Fig. \ref{sigma_1_norm}(c)), intercalating both peaks, cuts the frequency axis away from the origin, which also hints to a gapped, semimetallic behavior of the electronic structure in EuMnBi$_2$ \cite{Carbotte2016_2}. Overall, our comparative study broadly images the theoretical expectations within a minimum, simplified scenario, which however does not distinguish between type I and II Weyl semimetal \cite{Carbotte2016_2}. 

\begin{figure}[!htb]
\center
\includegraphics[width=7cm]{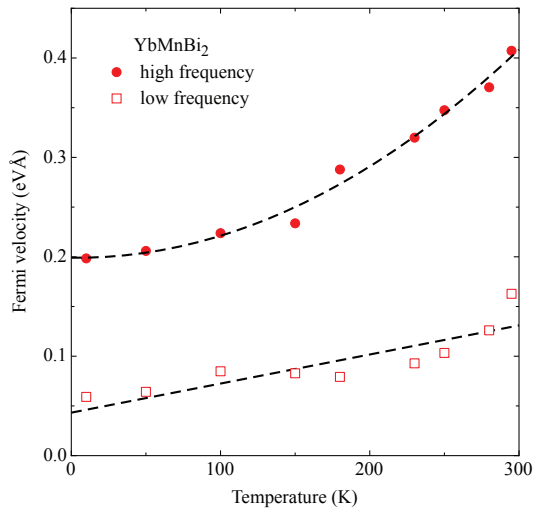}
\caption{(color online) Temperature dependence of the Fermi velocity ($v_F$) estimated from the linear frequency dependence of $\sigma_1(\omega)$ of YbMnBi$_2$ at low and high frequency intervals (see Fig. \ref{sigma_1_norm}). The dashed lines emphasize the squared and almost linear temperature dependence of $v_F$ at high and low frequencies, respectively.}
\label{Fermi_velocity}
\end{figure}

In principle, one can extract the Fermi velocity ($v_F$) from the slope of the linear optical conductivity as a function of frequency, as widely applied on several materials \cite{Chen2015,Sushkov2015,Timusk2013,Xu2016,Neubauer2016}, with the formula $\sigma_1(\omega) = \frac{NG_0\omega}{24v_F}$, $G_0$  being the quantum conductance and $N$ the number of Weyl nodes. We extract $v_F$ of YbMnBi$_2$ from the slope of $\sigma_1(\omega)$ in the low and high energy intervals, typically 30-90 and 200-400 cm$^{-1}$ respectively, where linearity is observed (Fig. \ref{sigma_1_norm}(a) and (c)). Figure \ref{Fermi_velocity} shows the results of our estimation as a function of temperature, considering $N$ = 8 \cite{Borisenko2016,comment2}. From ARPES we learn that the Weyl cones are very anisotropic with $v_F$ ranging between 0.043 and 9 eV\AA. Our optical experiment as a momentum-averaged probe does not allow any resolution in reciprocal space; in this respect our values for $v_F$ in both ranges and at all temperatures are reasonable, yet at the lower bound of the ARPES data \cite{Borisenko2016}. Our estimation may pick up the contribution by other interband transitions \cite{comment_bandstructure2}, not necessarily related to the Weyl cones, which could also explain the different $v_F$ in both ranges. Even though the estimation of $v_F$ is possible only on a reduced energy interval upon increasing the temperature \cite{comment}, its temperature dependence seems nonetheless quite intriguing, being linear in the low energy interval, despite the limited statistic due to the scattering of the data, and squared in the high one. Furthermore, $v_F$ at high energy scales doubles between 10 and 300 K. The temperature dependence of the chemical potential and of the band structure together with self-energy effects may considerably affect the temperature dependence of $v_F$ as well and it remains to be seen how this can be consistently reconciled within a theoretical framework.

In conclusion, our optical results clearly characterize YbMnBi$_2$ as a Weyl semimetal and EuMnBi$_2$ as its more conventional semimetal counterpart. Even though our findings compare reasonably well with the predicted optical fingerprints of generic Weyl semimetals \cite{Carbotte2016_2}, it is a challenge left for the future to develop a detailed description comprehensively taking into account the full complexity of the electronic properties and magnetism as well as of the specific type II Weyl semimetal nature of YbMnBi$_2$.

\section*{Acknowledgements}
The authors wish to thank S. Borisenko, J. Carbotte and M. Troyer for fruitful discussions. This work was supported by the Swiss National Science Foundation (SNSF). Work at Brookhaven National Laboratories was supported by the U.S. DOE-BES, Division of Materials Science and Engineering, under
Contract No. DE-SC0012704. L.D. acknowledges the hospitality at Aspen Center for Physics, where part of this paper was conceived. \\

$^+$ Both authors equally contributed to the experimental work.

$^\ast$ Correspondence and requests for materials should be addressed to: 
L. Degiorgi, Laboratorium f\"ur Festk\"orperphysik, ETH - Z\"urich, 8093 Z\"urich, Switzerland; 
email: degiorgi@solid.phys.ethz.ch



\begin{thebibliography}{10}

\bibitem{Richard2010} P. Richard et al., \emph{Phys. Rev. Lett.} \textbf{104}, 137001 (2010).

\bibitem{Novoselov2005} K.S. Novoselov et al., \emph{Nature} \textbf{438}, 197 (2005).

\bibitem{Kane2010} M.Z. Hasan and C.L. Kane, \emph{Rev. Mod. Phys.} \textbf{82}, 3045 (2010) and references therein.

\bibitem{Weyl1929} H. Weyl, \emph{Z. Phys.} \textbf{56}, 330 (1929).

\bibitem{Wan2011} X. Wan et al., \emph{Phys. Rev. B} \textbf{83}, 205101 (2011).

\bibitem{Xu2011} G. Xu et al., \emph{Phys. Rev. Lett.} \textbf{107}, 186806 (2011).

\bibitem{Burkov2011} A.A. Burkov et al., \emph{Phys. Rev. B} \textbf{84}, 235126 (2011).

\bibitem{Huang2015} S.-M. Huang et al., \emph{Nature Comm.} \textbf{6}, 7373 (2015).

\bibitem{Xu2015} S.-Y. Xu et al., \emph{Sci. Adv.} \textbf{1}, 10 (2015) (doi: 10.1126/sciadv.1501092).

\bibitem{Soluyanov2015} A.A. Soluyanov et al., \emph{Nature} \textbf{527}, 495 (2015).

\bibitem{Z_Wang2016} Z. Wang et al., \emph{Phys. Rev. Lett.} \textbf{117}, 056805 (2016).

\bibitem{Huang2016} L. Huang et al., \emph{Nat. Mat.} (2016) (doi: 10.1038/NMAT.4685).

\bibitem{Park2011} J. Park et al., \emph{Phys. Rev. Lett.} \textbf{107}, 126402 (2011).

\bibitem{Wang2011} K. Wang et al., \emph{Phys. Rev. B} \textbf{84}, 220401(R) (2011).

\bibitem{Wang2012} K. Wang et al., \emph{Phys. Rev. B} \textbf{85}, 041101 (2012).

\bibitem{May2014} A.F. May et al., \emph{Phys. Rev. B} \textbf{90}, 075109 (2014).

\bibitem{Lee2013} G. Lee et al., \emph{Phys. Rev. B} \textbf{87}, 245104 (2013).

\bibitem{Borisenko2016} S. Borisenko et al., \emph{arXiv:1507.04847}, unpublished (2015).
 
\bibitem{Chen2015} R.Y. Chen et al., \emph{Phys. Rev. B} \textbf{92}, 075107 (2015).

\bibitem{Sushkov2015} A.B. Sushkov et al., \emph{Phys. Rev. B} \textbf{92}, 241108(R) (2015).

\bibitem{Timusk2013} T. Timusk et al., \emph{Phys. Rev. B} \textbf{87}, 235121 (2013).

\bibitem{Xu2016} B. Xu et al., \emph{Phys. Rev. B} \textbf{93}, 121110(R) (2016).

\bibitem{Neubauer2016} D. Neubauer et al., \emph{Phys. Rev. B} \textbf{93}, 121202 (2016).

\bibitem{Ashby2014} P.E.C. Ashby and J.P. Carbotte, \emph{Phys. Rev. B} \textbf{89}, 245121 (2014).

\bibitem{Carbotte2016_1} C.J. Tabert et al., \emph{Phys. Rev. B} \textbf{93}, 085426 (2016).

\bibitem{Carbotte2016_2} C.J. Tabert and J.P. Carbotte, \emph{Phys. Rev. B} \textbf{93}, 085442 (2016).

\bibitem{Wang2016} A. Wang et al., \emph{arXiv:1604.01009}, unpublished (2016).

\bibitem{comment_Fermi_arc} The measurement of the optical reflectivity is a bulk sensitive technique, which prevents addressing the implications of the Fermi surface arcs, a typical surface state.

\bibitem{grunerbook} M. Dressel and G. Gr\"uner, {\itshape Electrodynamics of Solids}, Cambridge University Press, Cambridge, England (2002).

\bibitem{comment_bandstructure1} The resonance frequencies of the major absorptions in $\sigma_1(\omega)$ closely coincide with the direct transitions along the M-$\Gamma$-X directions in the reciprocal space, as predicted by band structure calculations supporting the ARPES data \cite{Borisenko2016}.

\bibitem{comment_magnetism} As already pointed out in Ref. \onlinecite{Borisenko2016}, the canted antiferromagnetic order serves the purpose to induce a time-reversal symmetry breaking, when performing band structure calculations compatible with the ARPES results. However, the real magnetic order needs still to be experimentally determined.

\bibitem{comment} Interestingly enough, by inspecting $\sigma_1(\omega)$ in Fig. \ref{sigma_1_norm}(a) we can convincingly state that both quasi-linear frequency dependences of the optical conductivity extrapolate to zero conductivity at the origin of the frequency axis at all temperatures.

\bibitem{comment2} A similar analysis for the Eu compound in the energy interval from 600 to 1500 cm$^{-1}$ within the assumption of $N$ = 8 leads to $v_F(T)$ values between 5.75 and 9.5 eV\AA, which gather around a linear temperature dependence.

\bibitem{comment_bandstructure2} For instance, the low frequency tails associated with the strong absorptions at 6000 and 10000 cm$^{-1}$ in the Yb and Eu compound, respectively, may definitely affect the slope of $\sigma_1(\omega)$ and consequently the estimation of $v_F$ in the high frequency interval at the least.
 
\end{thebibliography}
\end{document}